\documentstyle[11pt,aaspp4]{article}

\def\la{\mathrel{\hbox{\rlap{\hbox{\lower4pt\hbox{$\sim$}}}\hbox{$<$}}}}
\def\ga{\mathrel{\hbox{\rlap{\hbox{\lower4pt\hbox{$\sim$}}}\hbox{$>$}}}}
\def\etal{et al.\,\,}

\slugcomment{Accepted to {\it The Astrophysical Journal}}

\begin{document}

\title{A Bayesian Inference Analysis of the X-ray Cluster Luminosity-Temperature Relation}

\author{D. E. Reichart\altaffilmark{1,2}, F. J. Castander\altaffilmark{1}, R. C. Nichol\altaffilmark{2}}

\altaffiltext{1}{Department of Astronomy and Astrophysics, University of Chicago, 5640 South Ellis Avenue, Chicago, IL 60637} 
\altaffiltext{2}{Department of Physics, Carnegie Mellon University, 5000 Forbes Avenue, Pittsburgh, PA 15213}

\begin{abstract}

We present a Bayesian inference analysis of the Markevitch (1998) and Allen \& Fabian (1998) cooling flow corrected X-ray cluster temperature catalogs that constrains the slope and the evolution of the empirical X-ray cluster luminosity-temperature ($L$-$T$) relation.  
We find that for the luminosity range $10^{44.5}$ erg s$^{-1} \la L_{bol} \la 10^{46.5}$ erg s$^{-1}$ and the redshift range $z \la 0.5$, $L_{bol} \propto T^{2.80^{+0.15}_{-0.15}}(1+z)^{(0.91-1.12q_0)^{+0.54}_{-1.22}}$.  
We also determine the $L$-$T$ relation that one should use when fitting the Press-Schechter mass function to X-ray cluster luminosity catalogs such as the {\it Einstein} Medium Sensitivity Survey (EMSS) and the Southern Serendipitous High-Redshift Archival {\it ROSAT} Catalog (Southern SHARC), for which cooling flow corrected luminosities are not determined and a universal X-ray cluster temperature of $T = 6$ keV is assumed.  
In this case, $L_{bol} \propto T^{2.65^{+0.23}_{-0.20}}(1+z)^{(0.42-1.26q_0)^{+0.75}_{-0.83}}$ for the same luminosity and redshift ranges.

\end{abstract}

\keywords{cosmology: observations --- cosmology: theory --- galaxies: clusters: general --- galaxies: luminosity function, mass function --- X-rays: galaxies}

\section{Introduction}

Assuming that X-ray clusters correspond to virialized, dark matter halos, the Press-Schechter mass function (e.g., Press \& Schechter 1974; Lacey \& Cole 1993) describes how the X-ray-selected cluster mass function evolves with redshift and how the cosmological parameters affect this evolution.
In particular, the cosmological mass density parameter, $\Omega_m$, strongly affects how this mass function evolves above $M_\star$ ($\sim 10^{14}$ M$_{\sun}$).  Consequently, by fitting this mass function to present and future X-ray cluster catalogs, this cosmological parameter can be constrained.

Unfortunately, X-ray-selected cluster mass catalogs that span sufficiently broad ranges in $M$ and $z$ to constrain $\Omega_m$ do not yet exist.  
Since the Press-Schechter mass function already assumes that X-ray clusters are virialized, one may convert this mass function to a temperature function with the virial theorem; however, X-ray cluster temperature catalogs that span sufficiently broad ranges in $T$ and $z$ to strongly constrain $\Omega_m$ also do not yet exist (Viana \& Liddle 1998; Blanchard, Bartlett, \& Sadat 1998, however, see Henry 1997; Eke \etal 1998).  
However, several X-ray cluster luminosity catalogs span sufficiently broad ranges in $L$ and $z$ to strongly constrain $\Omega_m$, and the number of such catalogs is growing.  
However, to fit the Press-Schechter mass function to such catalogs, one must invoke a luminosity-temperature ($L$-$T$) relation in addition to the virial theorem.  Theoretically, a variety of $L$-$T$ relations have been proposed (e.g., Kaiser 1986; Evrard \& Henry 1991; Kaiser 1991); consequently, the $L$-$T$ relation should be determined empirically.
Until recently, the $L$-$T$ relations of temperature catalogs have suffered from much scatter (e.g., Edge \& Stewart 1991; David \etal 1994; Mushotzky \& Scharf 1997); however, recently, Markevitch (1998), Allen \& Fabian (1998), and Arnaud \& Evrard (1998) have published temperature catalogs with temperatures and luminosities that have either been corrected for, or avoided the effects of cooling flows (see \S2); the result is a significant reduction of this scatter.
This may be the key to determining cosmological parameters with X-ray cluster catalogs:  given a well-constrained $L$-$T$ relation, the Press-Schechter mass function may be fitted to the growing number of independent, high-redshift, high-luminosity X-ray cluster catalogs with well-understood selection functions (see \S 3).  
A well-defined $L$-$T$ relation is also necessary for the Press-Schechter mass function to be fitted to temperature and mass catalogs since all cluster catalogs with well-understood selection functions are X-ray selected; i.e., a well-constrained $L$-$T$ relation is necessary to accurately relate the selection function to the temperature or mass function.  For an example, see Henry (1997).

In this paper, we present a Bayesian inference analysis of the Markevitch (1998) and Allen \& Fabian (1998) cooling flow corrected temperature catalogs that constrains the slope and evolution of the empirical $L$-$T$ relation in the luminosity range $10^{44.5}$ erg s$^{-1} \la L_{bol} \la 10^{46.5}$ erg s$^{-1}$ and the redshift range $z \la 0.5$.
We also determine the $L$-$T$ relation that one should use when fitting the Press-Schechter mass function to luminosity catalogs for which cooling flow corrected luminosities are not determined and a universal X-ray cluster temperature of $T = 6$ keV, is assumed.
We do this in \S3.  We present the model and the data in \S2; we draw conclusions in \S4.

\section{The Model \& The Data}

Following the notation of Mathiesen \& Evrard (1998), we model X-ray clusters' bolometric luminosities with power laws in mass and redshift:
\begin{equation}
L_{bol} \propto M^p (1+z)^s.
\end{equation}
Combining equation (1) with the virial theorem yields the $L$-$T$ relation (e.g., Reichart \etal 1998):
\begin{equation}
L_{bol} \propto T^{\frac{3p}{2}} (1+z)^{s-\frac{3p}{2}}.
\end{equation}
The model of Cavaliere, Menci, \& Tozzi (1997) suggests that bolometric luminosity is not well-modeled by a power law in temperature over sufficiently broad temperature ranges:  $p$ varies from $p \sim 2$ for rich clusters to $p \sim 3$ for groups, in agreement with the observations of Edge \& Stewart (1991) and Ponman \etal (1996).  However, in this paper, we are only concerned with the values of $p$ and $s$ over the luminosity and redshift ranges that the temperature catalogs of Markevitch and Allen \& Fabian span:  $10^{44.5}$ erg s$^{-1} \la L_{bol} \la 10^{46.5}$ erg s$^{-1}$ and $z \la 0.5$.  Over these luminosity and redshift ranges, equations (1) and (2) are reasonable approximations.

Fabian \etal (1994) showed that cooling flows at the centers of X-ray clusters are responsible for most of the scatter in the empirical $L$-$T$ relation; this scatter is evident in the $L$-$T$ relations of the temperature catalogs of, e.g., Edge \& Stewart (1991), David \etal (1994), and Mushotzky \& Scharf (1997), the temperatures and luminosities of which are not corrected for the effects of cooling flows.  
However, the temperature catalogs of Markevitch (1998) and Allen \& Fabian (1998) are cooling flow corrected, resulting in a significant reduction of this scatter.  In Figure 1a, we plot the cooling flow contaminated temperatures and luminosities of Markevitch and Allen \& Fabian; in Figure 1b, we plot their cooling flow corrected measurements.

The temperature catalog of Markevitch spans the luminosity range $10^{44.5}$ erg s$^{-1} \la L_{bol} \la 10^{45.75}$ erg s$^{-1}$ and the redshift range $z \la 0.1$.  Cooling flow corrected temperatures and luminosities are measured in the same way for each X-ray cluster:  (1) cooling flow corrected temperatures are measured by modeling and then removing the cooling flow component of {\it ASCA} spectra of the central region of each X-ray cluster (Markevitch \etal 1998); and (2) cooling flow corrected luminosities are measured by excising the central region of {\it ROSAT} HRI images of each X-ray cluster, and then backfilling the excised region using a $\beta$-model.  Markevitch measures cooling flow corrected temperatures and luminosities for a total of 31 X-ray clusters.

The temperature catalog of Allen \& Fabian spans the luminosity range $10^{45.25}$ erg s$^{-1} \la L_{bol} \la 10^{46}$ erg s$^{-1}$ and the redshift range $z \la 0.5$.
Cooling flow corrected temperatures and luminosities are measured with two different models:  model A for non-cooling flow clusters and model C for cooling flow clusters.  Allen \& Fabian designate a cluster as a cooling flow cluster if the upper limit of its central cooling time, as measured from {\it ROSAT} HRI images, is less than $10^{10}$ years; otherwise they designate it as a non-cooling flow cluster.  
Corrected temperatures and luminosities are measured for the cooling flow clusters by modeling and then removing the cooling flow component of {\it ASCA} spectra of these clusters (Allen \& Fabian 1998; Allen \etal 1998); temperatures and luminosities of the non-cooling flow clusters are measured with an isothermal spectral model (Allen \& Fabian 1998; Allen \etal 1998). 
Allen \& Fabian measure corrected temperatures and luminosities for 21 cooling flow clusters and temperatures and luminosities for 9 non-cooling flow clusters.

\section{Bayesian Inference}

We do not simply fit equation (2) to the union of the Markevitch and Allen \& Fabian catalogs; instead, we reflect for a moment on how differences in how temperatures and luminosities are measured for these two catalogs, as well as for the two samples of Allen \& Fabian, can affect the results of such a fit.
For example, given that Markevitch and Allen \& Fabian do measure temperatures and luminosities differently, and in the case of luminosities, with different instruments, a small, average offset between their respective temperatures and/or luminosities would not be an unreasonable expectation.  However, the effect of such an offset can be significant:  since the Markevitch catalog is a low redshift sample and the Allen \& Fabian catalog is a higher redshift sample, a small offset between the $L$-$T$ relations of these two catalogs can significantly effect the value of $s-3p/2$, which measures how the $L$-$T$ relation evolves with redshift.
Also, since the Markevitch catalog is a lower luminosity sample than the Allen \& Fabian catalog, such an offset can also affect the value of $p$.
Furthermore, given that Allen \& Fabian measure temperatures and luminosities with different models for different clusters - model A for non-cooling flow clusters and model C for cooling flow clusters - a small offset between the $L$-$T$ relations of these two samples might not be an unreasonable expectation either.  In fact, Allen \& Fabian report such an offset between the $L$-$T$ relations of these two samples; however, as they note, this effect might also be do to physical differences between cooling flow and non-cooling flow clusters.

Equation (2) is a three parameter model; the parameters are $p$, $s$, and the proportionality factor, call it $L_0$.  To avoid biasing our results due to average offsets between the three samples - the Markevitch catalog, the model A sample of Allen \& Fabian, and the model C sample of Allen \& Fabian - we replace equation (2) with a five parameter model; the parameters are $p$, $s$, and three proportionality factors - one for each sample - $L_1$, $L_2$, and $L_3$.
Hence, the total $\chi^2$ is given by the sum of the $\chi^2$ of each of the three samples:
\begin{equation}
\chi^2(p,s,L_1,L_2,L_3) = \chi^2_1(p,s,L_1) + \chi^2_2(p,s,L_2) + \chi^2_3(p,s,L_3),
\end{equation}
where the subscript denotes from which sample the $\chi^2$ is computed.
By Bayes' theorem, the posterior probability distribution for $p$ and $s$, $P(p,s)$, is given by marginalizing the likelihood function, given by $e^{-\chi^2/2}$, over the other three parameters, assuming a flat prior probability distribution for all five parameters:
\begin{equation}
P(p,s) \propto \int_{L_1} \int_{L_2} \int_{L_3} e^{-\frac{1}{2}\chi^2(p,s,L_1,L_2,L_3)} dL_1 dL_2 dL_3.
\end{equation}
Given equation (3), equation (4) becomes:
\begin{equation}
P(p,s) \propto \left[\int_{L_1} e^{-\frac{1}{2}\chi^2_1(p,s,L_1)} dL_1\right] \left[\int_{L_2} e^{-\frac{1}{2}\chi^2_2(p,s,L_2)} dL_2\right] \left[\int_{L_3} e^{-\frac{1}{2}\chi^2_3(p,s,L_3)} dL_3\right].
\end{equation}
Hence, the posterior probability distribution is simply proportional to the product of the posterior probability distributions of each of the three samples:
\begin{equation}
P(p,s) \propto P_1(p,s) P_2(p,s) P_3(p,s).
\end{equation}

Before we compute these probability distributions, we address a final concern:  although by correcting their temperatures and luminosities for the effects of cooling flows, Markevitch and Allen \& Fabian significantly reduce the scatter in the empirical $L$-$T$ relation, they do not completely remove this scatter.  
Ignoring this scatter leads to sometimes incorrect and always overconstrained values of the fitted parameters.\footnote{Consider, for example, the frequentists' $\Delta \chi^2$ distribution:  $\Delta \chi^2(\sigma_{measured}^2) = \chi^2(\sigma_{measured}^2) - \chi^2_m(\sigma_{measured}^2) = \sum_i([(y_i-y)^2-(y_i-y_m)^2]/\sigma_{measured,i}^2) > \sum_i([(y_i-y)^2-(y_i-y_m)^2]/(\sigma_{measured,i}^2+\sigma_{intrinsic}^2)) = \chi^2(\sigma_{measured}^2+\sigma_{intrinsic}^2) - \chi^2_m(\sigma_{measured}^2+\sigma_{intrinsic}^2) = \Delta \chi^2(\sigma_{measured}^2+\sigma_{intrinsic}^2)$.  Hence, ignoring the intrinsic scatter of data about a model yields artificially high values of $\Delta \chi^2$, or in Bayesian terms, an artificially narrow likelihood function.}
Furthermore, Allen \& Fabian report that the scatter in their cooling flow cluster $L$-$T$ relation is greater than the scatter in their non-cooling flow cluster $L$-$T$ relation.  This suggests that either the non-cooling flow components of cooling flow clusters are physically more diverse than that of non-cooling flow clusters, or more likely, it is simply more difficult to model a cooling flow cluster than it is to model a non-cooling flow cluster.  
We deal with these issues by adding in quadrature to the 1 $\sigma$ error bars\footnote{We derive 1 $\sigma$ error bars by scaling the available 90\% error bars by a factor of 0.61.} in $\log{T}$ of each X-ray cluster in a given sample, a constant, $\sigma_{\log{T}}$, which measures the standard scatter in the $L$-$T$ relation of that sample.  Our measure of this standard scatter is $P(\chi^2|\nu) = 0.5$; a similar, yet somewhat cruder measure would be $\chi^2 = \nu$, where $\nu$ is the number of degrees of freedom.
We compute different values of $\sigma_{\log{T}}$ for each sample to deal with the observation of Allen \& Fabian that the $L$-$T$ relations of different samples have different scatters; consequently, we are not losing information from the lower scatter samples, nor biasing our results from the higher scatter samples, by adopting a single value of $\sigma_{\log{T}}$ that is indicative of all of the samples.
We list our values of $\sigma_{\log{T}}$ for each sample in Tables 1 \& 2.  We confirm that the scatter in the model C sample of Allen \& Fabian is greater than the scatter in the model A sample of Allen \& Fabian.

We now determine the posterior probability distributions, $P(p,s)$, of each of the three samples and combine them in accordance with equation (6).
Credible regions are determined by normalizing the posterior probability distributions (e.g., Gregory \& Loredo 1992).
In Figure 2, we plot the 1, 2, and 3 $\sigma$ credible regions of the posterior probability distributions of the 9 cluster, non-cooling flow sample (solid lines) and the 21 cluster, cooling flow sample (dotted lines) of Allen \& Fabian.
Although the cooling flow sample has more clusters, it is less constraining than the smaller, non-cooling flow sample, because its $L$-$T$ relation is more scattered than that of the non-cooling flow sample.
The straight line in this figure marks $L$-$T$ relations that do not evolve, given by $s - 3p/2 = 0$.
For the top panel of this figure, we used $q_0 = 0$ luminosities; for the bottom panel, we used $q_0 = 0.5$ luminosities.  
The dependence of these luminosities upon the value of the Hubble parameter is not important, since the Hubble parameter can be grouped with the proportionality factor of equation (2), which we marginalize over.
Finally, we determine one dimensional credible intervals for the parameters $L_1$, $p$, and $s$, as well as for the evolution parameter $s - 3p/2$, which we list in Tables 1 ($q_0 = 0$) \& 2 ($q_0 = 0.5$).

In Figure 3, we plot credible regions of the combined posterior probability distribution of the Allen \& Fabian catalog (solid lines) and the posterior probability distribution of the Markevitch catalog (dotted lines).  The combined posterior probability distribution of the Allen \& Fabian catalog is given by normalizing the product of the posterior probability distributions of their model A and model C samples (Figure 2), in accordance with equation (6).  Since the Markevitch catalog is a low redshift sample, it only weakly constrains the value of $s$, which is strongly coupled to the evolution parameter, $s - 3p/2$.  However, this catalog does place a useful constraint upon the value of $p$.  One dimensional credible intervals are again listed in Tables 1 \& 2.

In Figure 4, we plot the one dimensional posterior probability distributions, $P(p)$, $P(s)$, and $P(s-3p/2)$, of the combined posterior probability distribution, $P(p,s)$, of the Allen \& Fabian and Markevitch catalogs, which is determined in accordance with equation (6).  The dotted lines in this figure mark the 1, 2, and 3 $\sigma$ credible intervals; the 1 $\sigma$ credible intervals are also listed in Tables 1 \& 2.  The left panels are for $q_0 = 0$ and the right panels are for $q_0 = 0.5$.  The results are well-summarized by the following values:  $p = 1.86^{+0.10}_{-0.10}$ and $s = (3.77 - 1.26q_0)^{+0.48}_{-1.22}$, or $3p/2 = 2.80^{+0.15}_{-0.15}$ and $s-3p/2 = (0.91 - 1.12q_0)^{+0.54}_{-1.22}$.

However, these are not the equations that one wants to use when fitting the Press-Schechter mass function to X-ray cluster luminosity catalogs (\S1).  First of all, luminosity catalogs, such as the {\it Einstein} Medium Sensitivity Survey (EMSS) and the Southern Serendipitous High-Redshift Archival {\it ROSAT} Catalog (Southern SHARC), are not cooling flow corrected.  Secondly, since spectra are not measured for luminosity catalog clusters, photon count rates are only converted to fluxes and luminosities by assuming spectra for the X-ray clusters.  In the cases of these two catalogs, a $T = 6$ keV thermal bremsstrahlung spectrum is assumed for all of the X-ray clusters.
Consequently, the $L$-$T$ relation that one should use when fitting the Press-Schechter mass function to luminosity catalogs of this type is best determined by fitting equation (2) to cooling flow corrected temperatures - which better reflect the masses - and cooling flow contaminated, $T = 6$ keV luminosities - which better reflect the observations.  We derive such luminosities from the cooling flow contaminated luminosities of Markevitch and Allen \& Fabian by scaling their values to what they would have reported had they assumed a $T = 6$ keV thermal bremsstrahlung spectrum, given their respective bands.
We plot their cooling flow corrected temperatures and these cooling flow contaminated, $T = 6$ keV luminosities in Figure 1c.
Finally, we repeat the above analysis; the results are presented in Figures 5, 6, \& 7, and Tables 3 \& 4.  The results are well-summarized by the following values:  $p = 1.77^{+0.16}_{-0.13}$ and $s = (3.14 - 1.30q_0)^{+0.88}_{-0.86}$, or $3p/2 = 2.65^{+0.23}_{-0.20}$ and $s-3p/2 = (0.42 - 1.26q_0)^{+0.75}_{-0.83}$.

\section{Discussion \& Conclusions}

In the previous section, we found that $s$ depends upon $q_0$, and we assumed that this dependence upon $q_0$ is linear.  Furthermore, we found that $p$ does not depend upon $q_0$.  These results are easily verified analytically.  Let subscript $q_0$ denote ``for an arbitrary value of $q_0$'', and let subscript zero denote ``for $q_0 = 0$''.  Then, 
\begin{equation}
L_{q_0} = L_0\left(\frac{d_{q_0}}{d_0}\right)^2,
\end{equation}
where $L$ is bolometric luminosity and $d$ is luminosity distance.
To first order in $z$, equation (7) is equivalent to:
\begin{equation}
L_{q_0} = L_0(1+z)^{-q_0}.
\end{equation}
Together, equations (2) and (8) imply that
\begin{equation}
L_{q_0} \propto T^{\frac{3p_0}{2}} (1+z)^{s_0-q_0-\frac{3p_0}{2}}.
\end{equation}
Hence, also by equation (2), $p_{q_0} \approx p_0$ and $s_{q_0} \approx s_0 - q_0$.
This verifies both that the dependence of $s$ upon $q_0$ is linear, and that the magnitude of this dependence is about unity.  This also verifies that $p$ is independent of $q_0$.

Arnaud \& Evrard (1998) measure the value of $3p/2$ from a sample of 24 non-cooling flow and weak cooling flow clusters.  They measure $3p/2 = 2.88 \pm 0.15$, which is in excellent agreement with our value:  $3p/2 = 2.80^{+0.15}_{-0.15}$.  Since their sample is assembled from 18 sources from the literature, we felt that it would be too difficult to deal with potential biases between subsamples of their catalog, as we did in this paper between the three samples of Markevitch and Allen \& Fabian; consequently, we did not include their catalog in our analysis.  However, the fact that our results are in such excellent agreement is reassuring.

In conclusion, we have constrained the slope and the evolution of the empirical $L$-$T$ relation using the cooling flow corrected X-ray cluster temperature catalogs of Markevitch and Allen \& Fabian, and Bayesian inference.  For the luminosity and redshift ranges $10^{44.5}$ erg s$^{-1} \la L_{bol} \la 10 ^{46.5}$ erg s$^{-1}$ and $z \la 0.5$, we find that $L_{bol} \propto T^{2.80^{+0.15}_{-0.15}}(1+z)^{(0.91-1.12q_0)^{+0.54}_{-1.22}}$.  
Hence, we find that the $L$-$T$ relation is consistent with no evolution over this redshift range; however, we also find that the evolution parameter, $s - 3p/2$, is a function of $q_0$.
We have also determined the $L$-$T$ relation that one should use when fitting the Press-Schechter mass function to X-ray cluster luminosity catalogs such as the EMSS and the Southern SHARC.  It differs from the above $L$-$T$ relation for the reasons stated in \S3. 
Given the growing number of independent, high-redshift, high-luminosity X-ray cluster catalogs with well-understood selection functions, a well-constrained $L$-$T$ relation may be the key to measuring cosmological parameters with X-ray cluster catalogs.

\acknowledgments
This research has been partially funded by NASA grants NAG5-6548 and NAG5-2432.  We are very grateful to C. Graziani and D. Q. Lamb for enlightening discussions about Bayesian inference.
D. E. R. is especially grateful to Dr. and Mrs. Bernard Keisler for their hospitality during the summer of 1998.  

\clearpage

\begin{deluxetable}{cccccccccc}
\footnotesize
\tablecolumns{5}
\tablewidth{0pc}
\tablecaption{The $L$-$T$ Relation:  $q_0 = 0$}
\tablehead{\colhead{$L_{bol}$ Range\tablenotemark{a}} & \colhead{$z$ Range} & \colhead{CF/NCF\tablenotemark{b}} & \colhead{$\log{L_i}$\tablenotemark{c}} & \colhead{$p$} & \colhead{$s$} & \colhead{$3p/2$} & \colhead{$s-3p/2$} & \colhead{$\sigma_{\log{T}}$} & \colhead{Catalog(s)\tablenotemark{d}}}
\startdata
$10^{45.25} - 10^{46.5}$ & $\la$ 0.5 & NCF & $45.42^{+0.10}_{-0.10}$ & $2.06^{+0.31}_{-0.26}$ & $3.38^{+1.17}_{-1.13}$ & $3.09^{+0.47}_{-0.39}$ & $0.38^{+1.12}_{-1.33}$ & $0.024^{+0.013}_{-0.009}$ & 1 \nl   
$10^{45.25} - 10^{46.5}$ & $\la$ 0.5 & CF & $45.70^{+0.14}_{-0.11}$ & $1.82^{+0.65}_{-0.43}$ & $2.21^{+2.22}_{-1.16}$ & $2.74^{+0.97}_{-0.64}$ & $-0.35^{+1.84}_{-1.66}$ & $0.054^{+0.028}_{-0.021}$ & 1 \nl   
$10^{45.25} - 10^{46.5}$ & $\la$ 0.5 & CF+NCF & $-$ & $1.98^{+0.25}_{-0.20}$ & $2.79^{+1.26}_{-0.59}$ & $2.96^{+0.37}_{-0.30}$ & $-0.10^{+1.20}_{-0.74}$ & $-$ & 1 \nl
$10^{44.5} - 10^{45.75}$ & $\la$ 0.1 & CF+NCF & $45.29^{+0.10}_{-0.09}$ & $1.76^{+0.12}_{-0.11}$ & $8.55^{+3.39}_{-3.58}$ & $2.64^{+0.18}_{-0.17}$ & $6.01^{+3.28}_{-3.67}$ & $0.034^{+0.007}_{-0.006}$ & 2 \nl 
$10^{44.5} - 10^{46.5}$ & $\la$ 0.5 & CF+NCF & $-$ & $1.86^{+0.10}_{-0.10}$ & $3.77^{+0.48}_{-1.26}$ & $2.80^{+0.15}_{-0.15}$ & $0.91^{+0.55}_{-1.19}$ & $-$ & 1+2 \nl
\enddata
\tablenotetext{a}{erg s$^{-1}$.}
\tablenotetext{b}{Cooling flow / Non-cooling flow.}
\tablenotetext{c}{For $T$ measured in units of 8 keV.}
\tablenotetext{d}{1. Allen \& Fabian 1998; 2. Markevitch 1998.}
\end{deluxetable}

\clearpage

\begin{deluxetable}{cccccccccc}
\footnotesize
\tablecolumns{5}
\tablewidth{0pc}
\tablecaption{The $L$-$T$ Relation:  $q_0 = 0.5$}
\tablehead{\colhead{$L_{bol}$ Range\tablenotemark{a}} & \colhead{$z$ Range} & \colhead{CF/NCF\tablenotemark{b}} & \colhead{$\log{L_i}$\tablenotemark{c}} & \colhead{$p$} & \colhead{$s$} & \colhead{$3p/2$} & \colhead{$s-3p/2$} & \colhead{$\sigma_{\log{T}}$} & \colhead{Catalog(s)\tablenotemark{c}}}
\startdata
$10^{45.25} - 10^{46.5}$ & $\la 0.5$ & NCF & $45.42^{+0.10}_{-0.10}$ & $2.06^{+0.31}_{-0.26}$ & $2.87^{+1.15}_{-1.27}$ & $3.09^{+0.47}_{-0.39}$ & $-0.25^{+1.12}_{-1.27}$ & $0.024^{+0.013}_{-0.009}$ & 1 \nl
$10^{45.25} - 10^{46.5}$ & $\la 0.5$ & CF & $45.72^{+0.12}_{-0.12}$ & $1.82^{+0.65}_{-0.43}$ & $1.60^{+2.23}_{-1.15}$ & $2.74^{+0.97}_{-0.64}$ & $-0.84^{+1.69}_{-1.84}$ & $0.054^{+0.028}_{-0.021}$ & 1 \nl            
$10^{45.25} - 10^{46.5}$ & $\la 0.5$ & CF+NCF & $-$ & $1.98^{+0.25}_{-0.20}$ & $2.15^{+1.30}_{-0.56}$ & $2.96^{+0.37}_{-0.30}$ & $-0.68^{+1.20}_{-0.79}$ & $-$ & 1 \nl
$10^{44.5} - 10^{45.75}$ & $\la 0.1$ & CF+NCF & $45.29^{+0.10}_{-0.09}$ & $1.76^{+0.12}_{-0.11}$ & $8.05^{+3.32}_{-3.58}$ & $2.64^{+0.18}_{-0.17}$ & $5.18^{+3.60}_{-3.42}$ & $0.034^{+0.007}_{-0.006}$ & 2 \nl
$10^{44.5} - 10^{46.5}$ & $\la 0.5$ & CF+NCF & $-$ & $1.86^{+0.10}_{-0.10}$ & $3.14^{+0.49}_{-1.19}$ & $2.80^{+0.15}_{-0.15}$ & $0.35^{+0.52}_{-1.26}$ & $-$ & 1+2 \nl
\enddata
\tablenotetext{a}{erg s$^{-1}$.}
\tablenotetext{b}{Cooling flow / Non-cooling flow.}
\tablenotetext{c}{For $T$ measured in units of 8 keV.}
\tablenotetext{d}{1. Allen \& Fabian 1998; 2. Markevitch 1998.}
\end{deluxetable}

\clearpage

\begin{deluxetable}{cccccccccc}
\footnotesize
\tablecolumns{5}
\tablewidth{0pc}
\tablecaption{The $L$-$T$ Relation For Fitting The Press-Schechter Mass Function To X-ray Cluster Luminosity Catalogs:  $q_0 = 0$}
\tablehead{\colhead{$L_{bol}$ Range\tablenotemark{a}} & \colhead{$z$ Range} & \colhead{CF/NCF\tablenotemark{b}} & \colhead{$\log{L_i}$\tablenotemark{c}} & \colhead{$p$} & \colhead{$s$} & \colhead{$3p/2$} & \colhead{$s-3p/2$} & \colhead{$\sigma_{\log{T}}$} & \colhead{Catalog(s)\tablenotemark{c}}}
\startdata
$10^{45.25} - 10^{46.5}$ & $\la 0.5$ & NCF & $45.42^{+0.10}_{-0.09}$ & $1.91^{+0.29}_{-0.23}$ & $3.25^{+1.08}_{-1.05}$ & $2.86^{+0.44}_{-0.34}$ & $0.39^{+1.06}_{-1.12}$ & $0.023^{+0.013}_{-0.009}$ & 1 \nl   
$10^{45.25} - 10^{46.5}$ & $\la 0.5$ & CF & $45.68^{+0.11}_{-0.11}$ & $1.76^{+0.55}_{-0.29}$ & $2.26^{+1.67}_{-1.32}$ & $2.65^{+0.82}_{-0.43}$ & $-0.15^{+1.35}_{-1.80}$ & $0.048^{+0.028}_{-0.020}$ & 1 \nl   
$10^{45.25} - 10^{46.5}$ & $\la 0.5$ & CF+NCF & $-$ & $1.88^{+0.22}_{-0.19}$ & $3.04^{+0.73}_{-0.85}$ & $2.82^{+0.32}_{-0.29}$ & $0.14^{+0.92}_{-0.87}$ & $-$ & 1 \nl
$10^{44.5} - 10^{45.75}$ & $\la 0.1$ & CF+NCF & $45.26^{+0.18}_{-0.15}$ & $1.52^{+0.25}_{-0.19}$ & $9.22^{+5.04}_{-5.63}$ & $2.28^{+0.37}_{-0.28}$ & $6.82^{+5.29}_{-5.72}$ & $0.065^{+0.011}_{-0.009}$ & 2 \nl
$10^{44.5} - 10^{46.5}$ & $\la 0.5$ & CF+NCF & $-$ & $1.77^{+0.15}_{-0.13}$ & $3.14^{+0.65}_{-0.88}$ & $2.65^{+0.23}_{-0.20}$ & $0.42^{+0.73}_{-0.84}$ & $-$ & 1+2 \nl
\enddata
\tablenotetext{a}{erg s$^{-1}$.}
\tablenotetext{b}{Cooling flow / Non-cooling flow.}
\tablenotetext{c}{For $T$ measured in units of 8 keV.}
\tablenotetext{d}{1. Allen \& Fabian 1998; 2. Markevitch 1998.}
\end{deluxetable}

\clearpage

\begin{deluxetable}{cccccccccc}
\footnotesize
\tablecolumns{5}
\tablewidth{0pc}
\tablecaption{The $L$-$T$ Relation For Fitting The Press-Schechter Mass Function To X-ray Cluster Luminosity Catalogs:  $q_0 = 0.5$}
\tablehead{\colhead{$L_{bol}$ Range\tablenotemark{a}} & \colhead{$z$ Range} & \colhead{CF/NCF\tablenotemark{b}} & \colhead{$\log{L_i}$\tablenotemark{c}} & \colhead{$p$} & \colhead{$s$} & \colhead{$3p/2$} & \colhead{$s-3p/2$} & \colhead{$\sigma_{\log{T}}$} & \colhead{Catalog(s)\tablenotemark{c}}}
\startdata
$10^{45.25} - 10^{46.5}$ & $\la 0.5$ & NCF & $45.44^{+0.09}_{-0.10}$ & $1.91^{+0.29}_{-0.23}$ & $2.62^{+1.08}_{-0.99}$ & $2.86^{+0.44}_{-0.34}$ & $-0.23^{+1.15}_{-1.18}$ & $0.023^{+0.013}_{-0.009}$ & 1 \nl
$10^{45.25} - 10^{46.5}$ & $\la 0.5$ & CF & $45.68^{+0.11}_{-0.10}$ & $1.76^{+0.55}_{-0.29}$ & $1.65^{+1.65}_{-1.28}$ & $2.65^{+0.82}_{-0.43}$ & $-0.82^{+1.38}_{-1.69}$ & $0.048^{+0.028}_{-0.020}$ & 1 \nl
$10^{45.25} - 10^{46.5}$ & $\la 0.5$ & CF+NCF & $-$ & $1.88^{+0.22}_{-0.19}$ & $2.41^{+0.81}_{-0.86}$ & $2.82^{+0.33}_{-0.29}$ & $-0.43^{+0.79}_{-0.80}$ & $-$ & 1 \nl
$10^{44.5} - 10^{45.75}$ & $\la 0.1$ & CF+NCF & $45.26^{+0.18}_{-0.15}$ & $1.52^{+0.25}_{-0.19}$ & $8.72^{+5.04}_{-5.71}$ & $2.28^{+0.37}_{-0.28}$ & $6.32^{+5.23}_{-5.72}$ & $0.065^{+0.011}_{-0.009}$ & 2 \nl
$10^{44.5} - 10^{46.5}$ & $\la 0.5$ & CF+NCF & $-$ & $1.77^{+0.16}_{-0.13}$ & $2.49^{+0.71}_{-0.84}$ & $2.65^{+0.23}_{-0.20}$ & $-0.21^{+0.77}_{-0.82}$ & $-$ & 1+2 \nl
\enddata
\tablenotetext{a}{erg s$^{-1}$.}
\tablenotetext{b}{Cooling flow / Non-cooling flow.}
\tablenotetext{c}{For $T$ measured in units of 8 keV.}
\tablenotetext{d}{1. Allen \& Fabian 1998; 2. Markevitch 1998.}
\end{deluxetable}

\clearpage

\figcaption[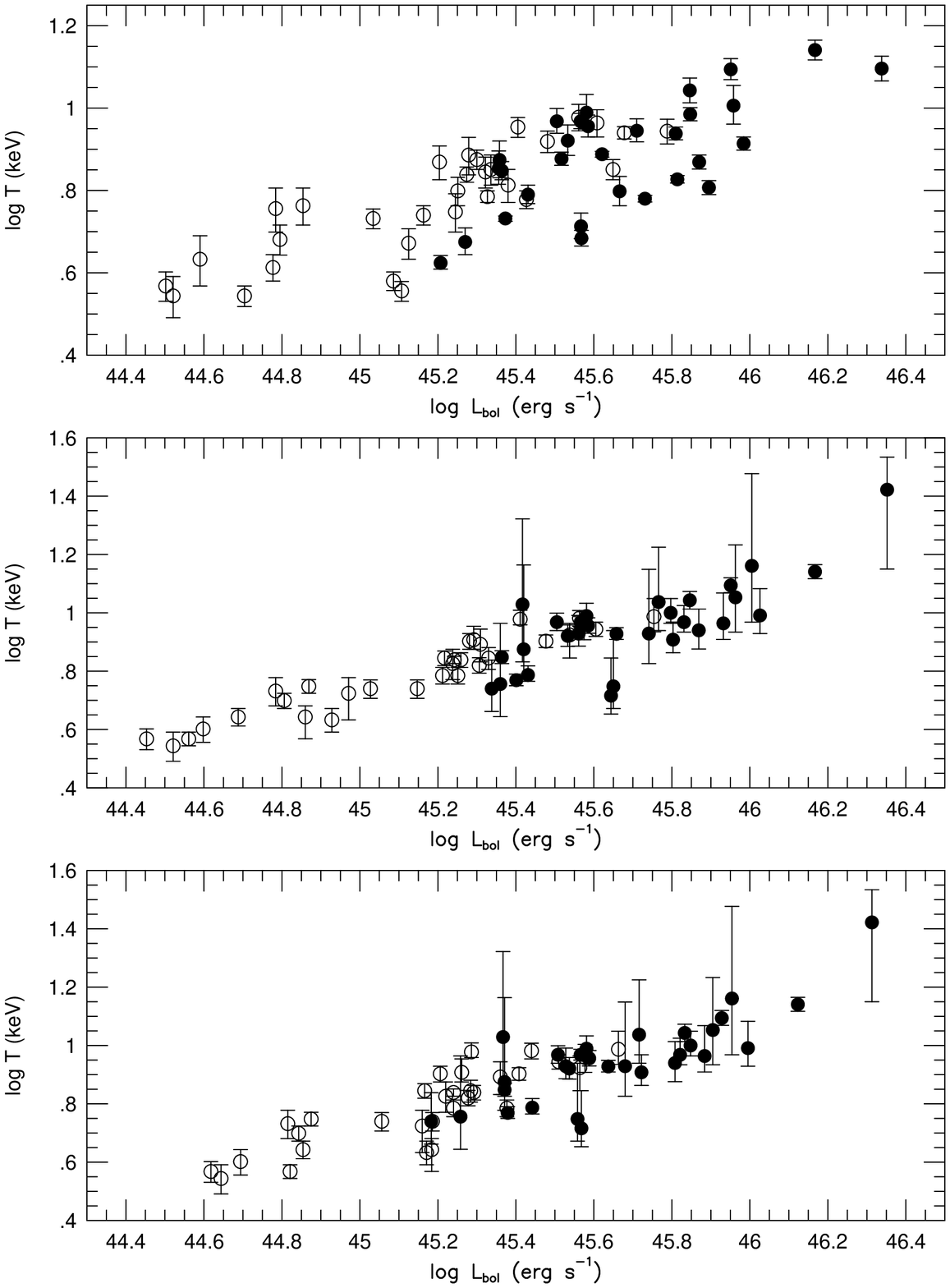]{The X-ray cluster temperature catalogs of Markevitch (open circles) and Allen \& Fabian (solid circles).  Cooling flow contaminated temperatures and luminosities are plotted in the top panel (Figure 1a); cooling flow corrected temperatures and luminosities are plotted in the middle panel (Figure 1b).  Cooling flow corrected temperatures versus cooling flow uncorrected, $T = 6$ keV luminosities are plotted in the bottom panel (Figure 1c) (see \S3).  Error bars are 90\% confidence intervals.\label{ltfig1.ps}}

\figcaption[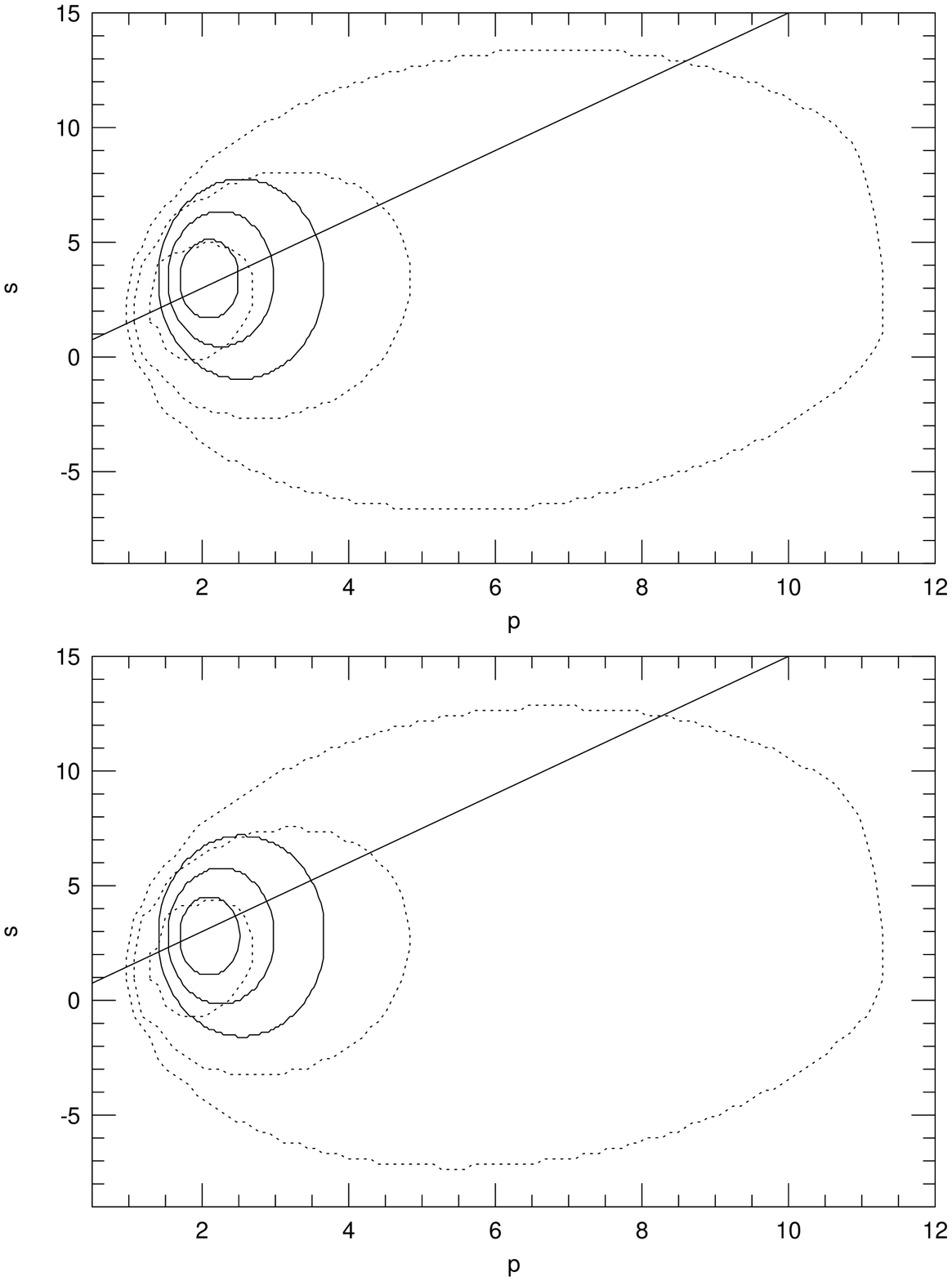]{The 1, 2, and 3 $\sigma$ credible regions of the posterior probability distributions of the non-cooling flow sample (solid lines) and the cooling flow sample (dotted lines) of Allen \& Fabian (see \S3).  The straight line in this figure marks $L$-$T$ relations that do not evolve, given by $s - 3p/2 = 0$.  The top panel is for $q_0 = 0$; the bottom panel is for $q_0 = 0.5$.\label{ltfig2.ps}}

\figcaption[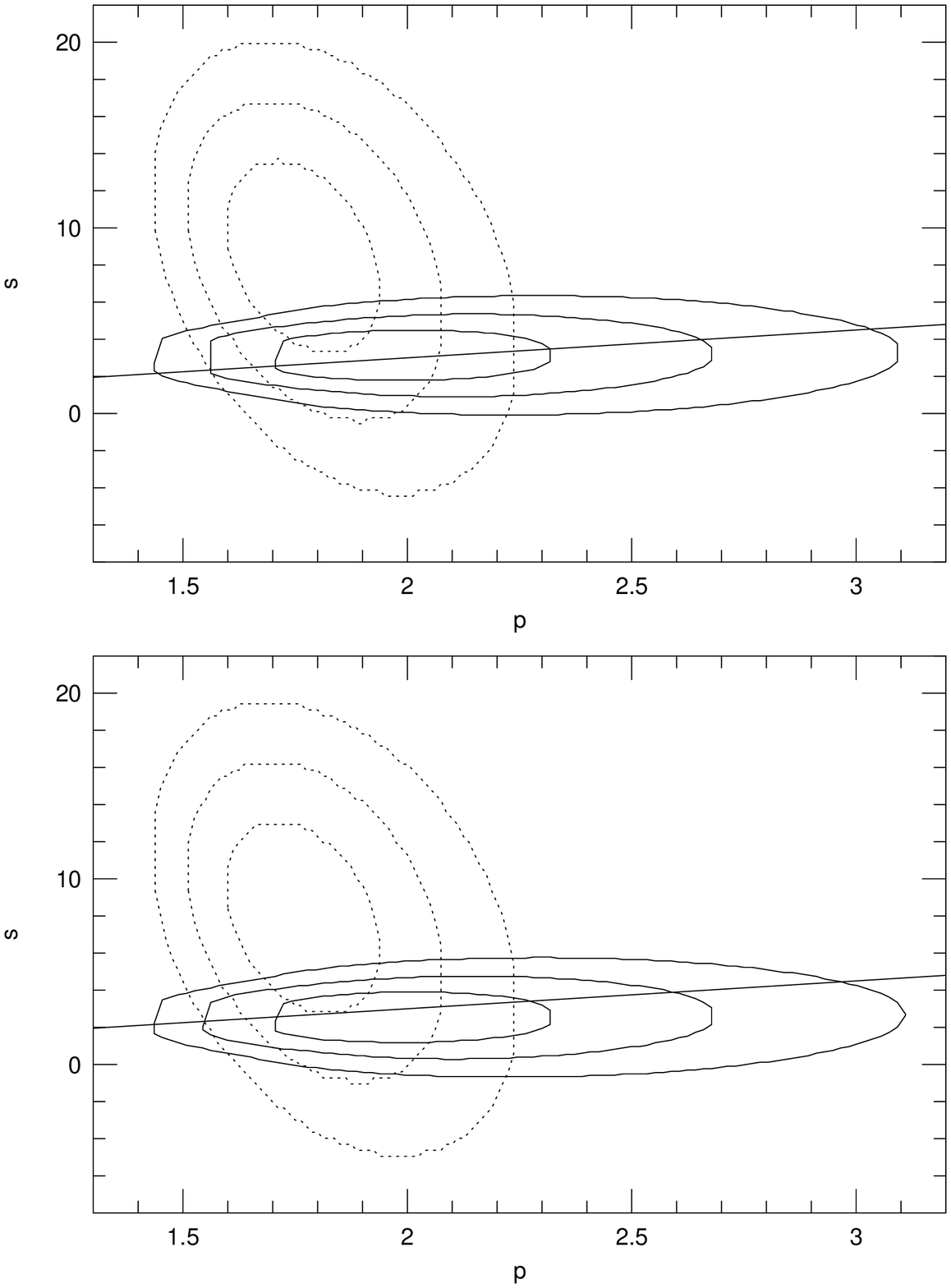]{The same as Figure 2, except for the combined posterior probability distribution of the Allen \& Fabian catalog (solid lines) and the posterior probability distribution of the Markevitch catalog (dotted lines) (see \S3).\label{ltfig3.ps}}

\figcaption[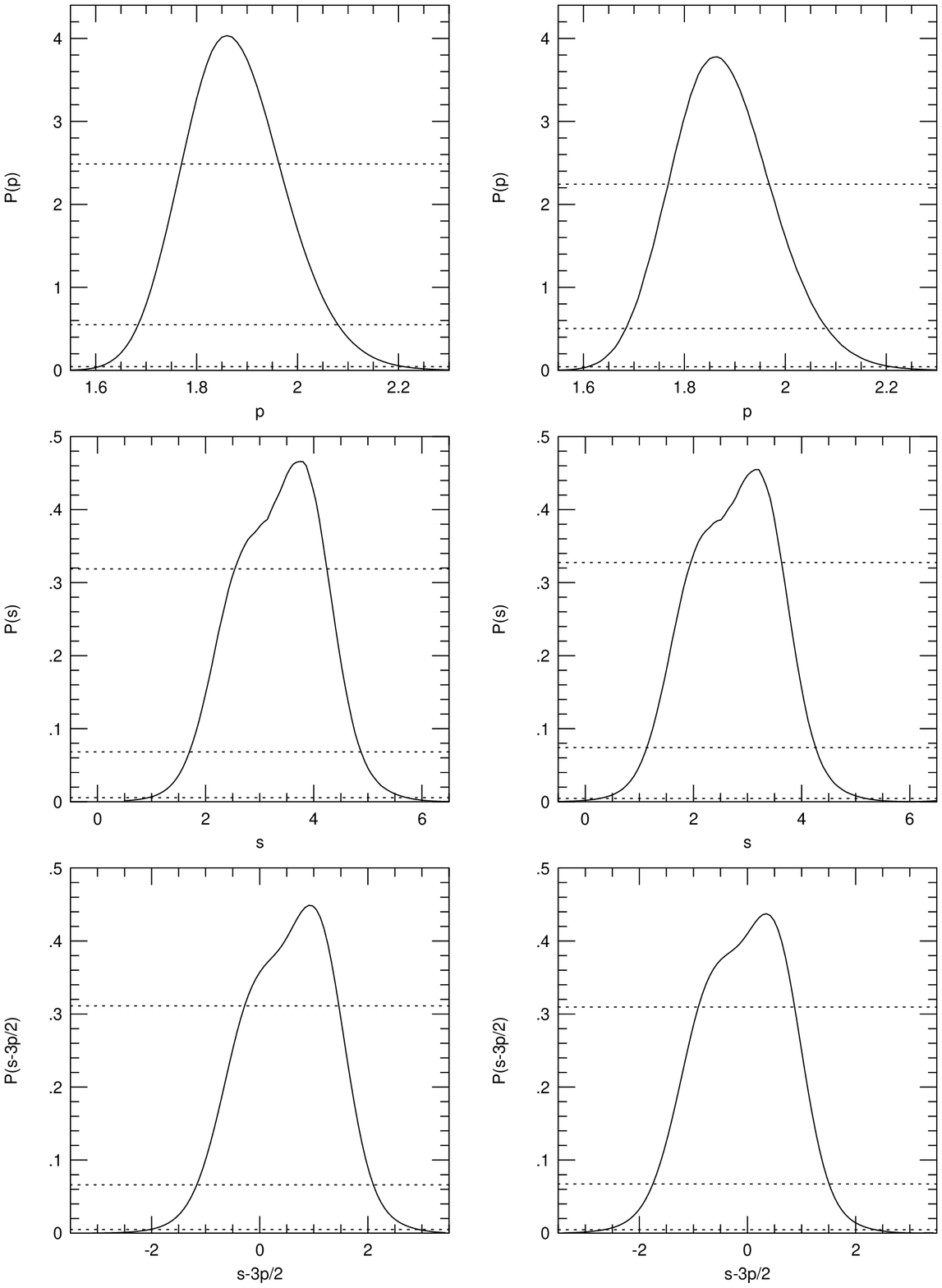]{The one dimensional posterior probability distributions, $P(p)$, $P(s)$, and $P(s-3p/2)$, of the combined posterior probability distribution, $P(p,s)$, of the Allen \& Fabian and Markevitch catalogs (see \S3).  The dotted lines mark the 1, 2, and 3 $\sigma$ credible intervals.  The left panels are for $q_0 = 0$ and the right panels are for $q_0 = 0.5$.\label{ltfig4.ps}}

\figcaption[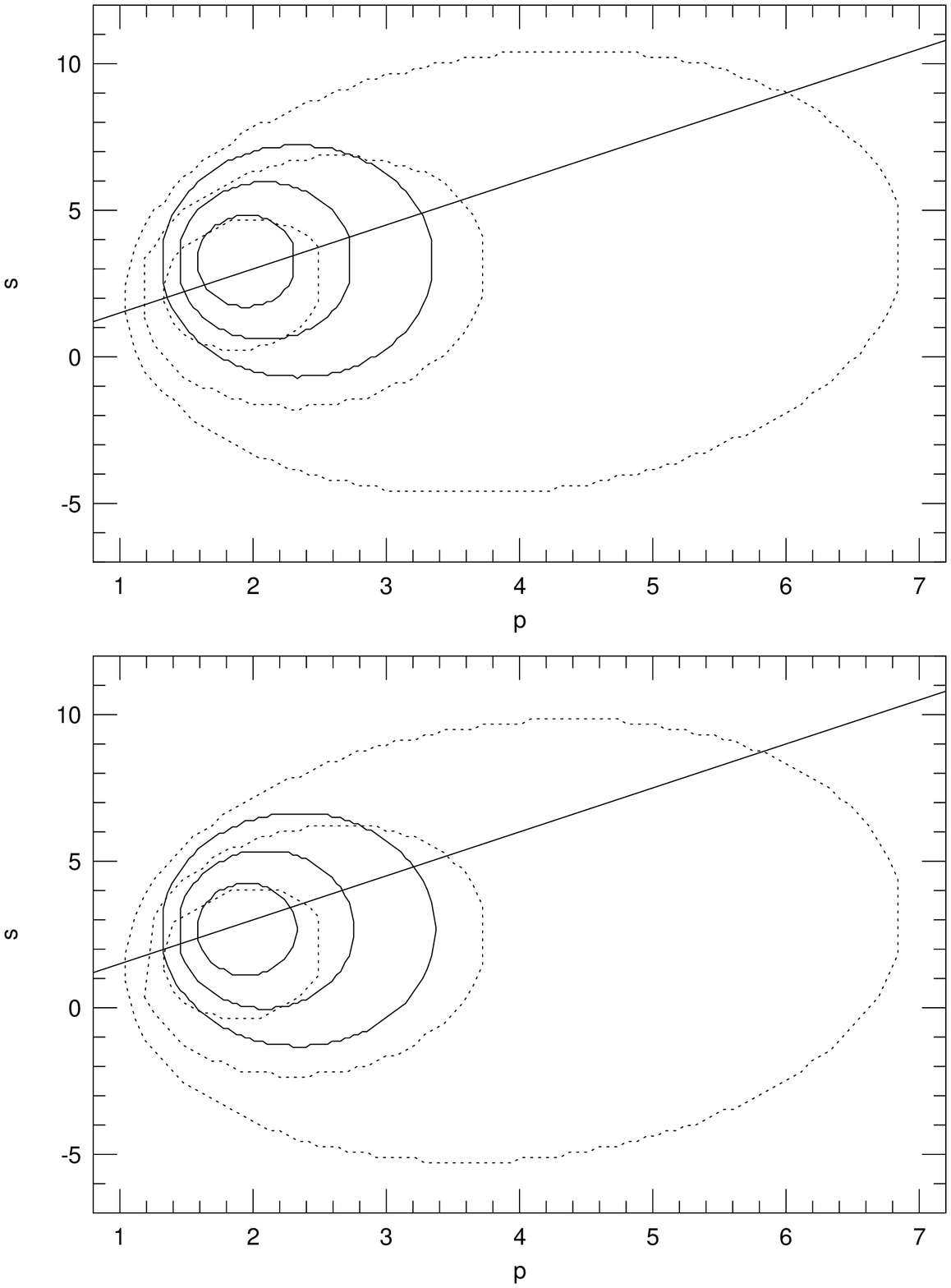]{The same as Figure 2, except with cooling flow contaminated, $T = 6$ keV luminosities (see \S3).\label{ltfig5.ps}}

\figcaption[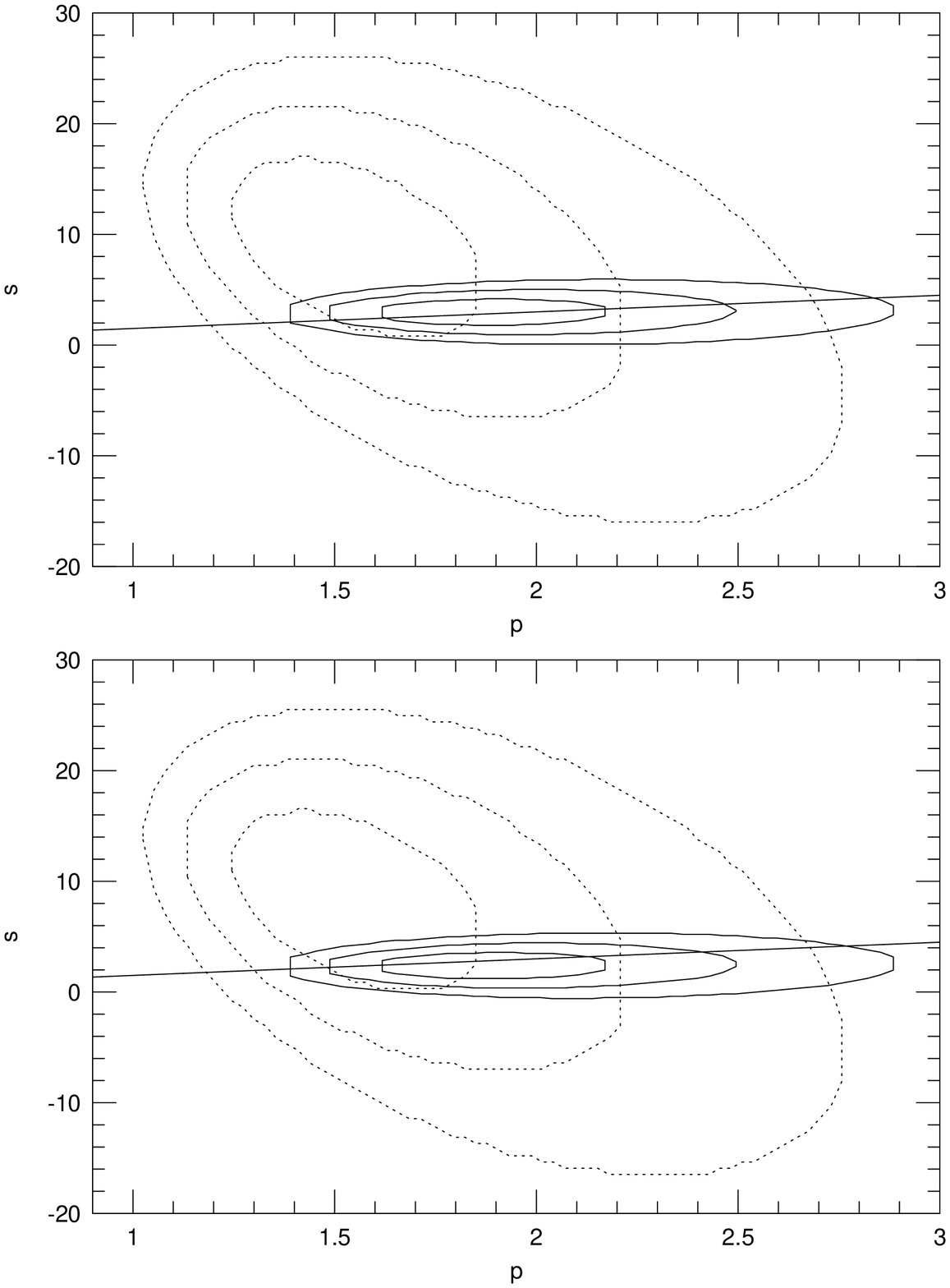]{The same as Figure 3, except with cooling flow contaminated, $T = 6$ keV luminosities (see \S3).\label{ltfig6.ps}}

\figcaption[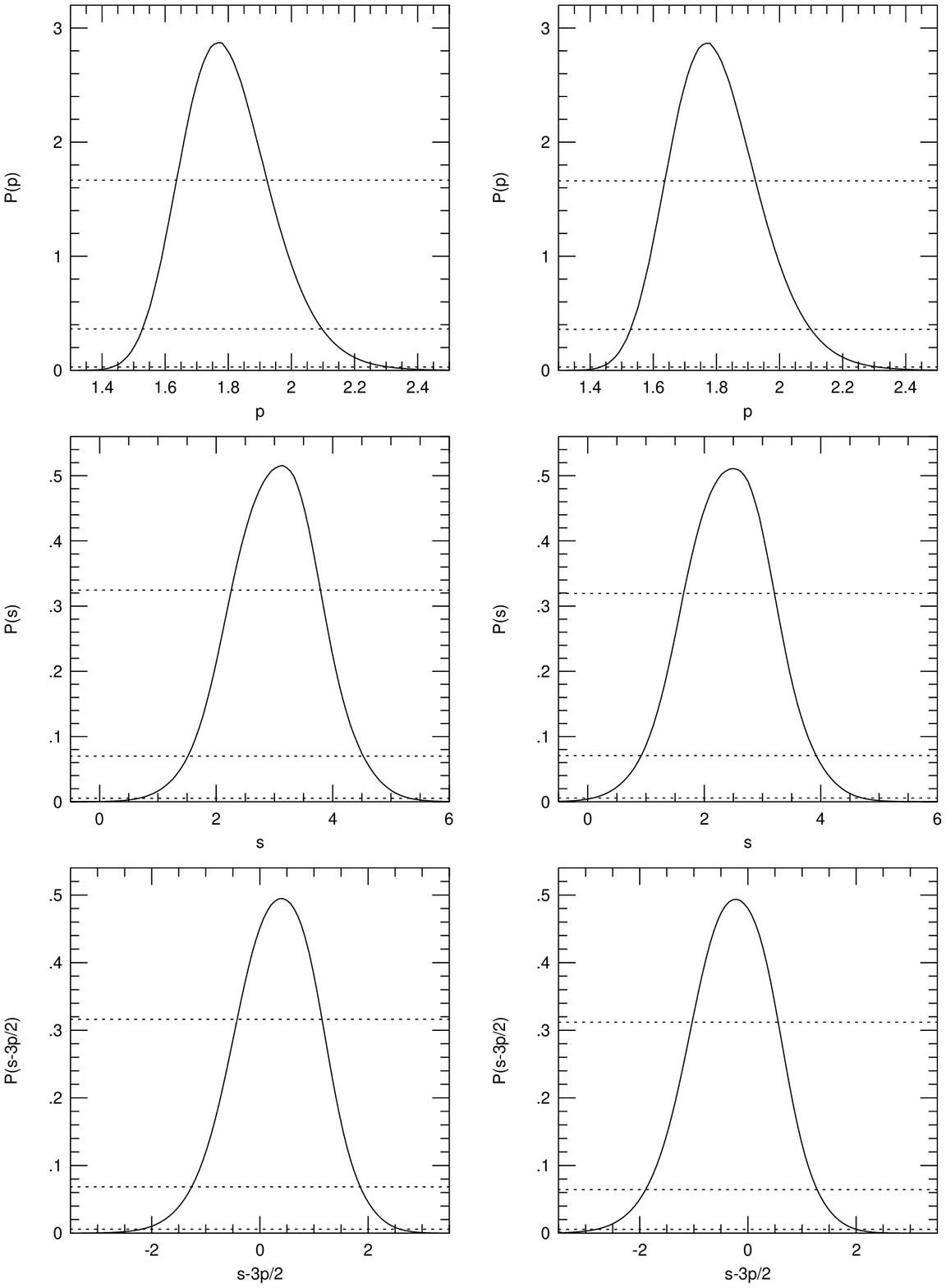]{The same as Figure 4, except with cooling flow contaminated, $T = 6$ keV luminosities (see \S3).\label{ltfig7.ps}}

\clearpage

\setcounter{figure}{0}

\begin{figure}[tb]
\plotone{ltfig1.ps}
\end{figure}

\begin{figure}[tb]
\plotone{ltfig2.ps}
\end{figure}

\begin{figure}[tb]
\plotone{ltfig3.ps}
\end{figure}

\begin{figure}[tb]
\plotone{ltfig4.ps}
\end{figure}

\begin{figure}[tb]
\plotone{ltfig5.ps}
\end{figure}

\begin{figure}[tb]
\plotone{ltfig6.ps}
\end{figure}

\begin{figure}[tb]
\plotone{ltfig7.ps}
\end{figure}

\end{document}